\documentclass{aa}  

\usepackage{txfonts}
\usepackage{multirow,graphicx}
\DeclareMathAlphabet{\mathsfit}{T1}{\sfdefault}{\mddefault}{\sldefault}
\usepackage[dvipsnames]{xcolor}
\begin{document}

   \title{Modelling of the scandium abundance evolution in AmFm stars}


   \author{A. Hui-Bon-Hoa\inst{1},
          G. Alecian\inst{2}
                    \and
          F. LeBlanc\inst{3}
          }

   \institute{IRAP, Universit\'e de Toulouse, CNRS, UPS, CNES, Toulouse, France\\
            \email{alain.hui-bon-hoa@irap.omp.eu}
   \and
   LUTH, Observatoire de Paris, PSL Research University, CNRS, Université Paris Diderot, 5 place Jules Janssen, F-92190 Meudon, France
   \and
   Département de Physique et d'Astronomie, Université de Moncton, Moncton, NB, E1A 3E9, Canada\\
             }

   \date{Received 16 June 2022; accepted 18 September 2022}

  \abstract
    {Scandium is a key element of the Am star phenomenon since its surface under-abundance is one of the criteria that characterise such stars. Thanks to the availability of a sufficiently complete set of theoretical atomic data for this element, reliable radiative accelerations for Sc can now be computed, which allows its behaviour under the action of atomic diffusion to be modelled.}
   {We explore the required conditions, in terms of mixing processes or mass loss, for our models to reproduce the observed surface abundances of Sc in Am stars.}
   {The models are computed with the Toulouse-Geneva evolution code, which uses the parametric single-valued parameter method for the calculation of radiative accelerations. Fingering mixing is included, using a prescription that comes from 3D hydrodynamical simulations. Other parameter-dependent turbulent mixing processes are also considered. A global mass loss is also implemented.}
   {When no mass loss is considered, the observed abundances of Sc are rather in favour of the models whose superficial layers are fully mixed down to the iron accumulation zone, although other mixing prescriptions are also able to reproduce the observations for the most massive model presented here (2.0~${M_\odot}$). The models including mass loss with rates in the range of $[10^{-13};10^{-14}]~{M_\odot/\mathrm {yr}}$ are compatible with some of the observations, while other observations suggest that the mass-loss rate could be lower. The constraints brought by the modelling of Sc are consistent with those derived using other chemical elements.}
   {}


   \maketitle
%

\section{Introduction}

The Am or metallic-line stars were identified as a group by \cite{Titus_Morgan1940} while classifying the members of the Hyades open cluster. They remarked that a subset of stars had weaker calcium lines than expected relative to the hydrogen line strengths, whereas the metallic lines were stronger. In the higher-dispersion spectra used in abundance determinations, scandium lines were also weaker \citep[e.g.][]{Jaschek_Jaschek1960}. In his review, \cite{Conti1970} concluded that 'Am star atmospheres are characterized by either deficient Ca (Sc) or over-abundant heavier elements, or both', which put forward surface abundance anomalies as the main characterisation criterion for an Am star. The same kind of chemical peculiarities has been observed in F-type stars \citep[e.g.][]{Cowley1976}, and so one often refers to the metallic-line stars as AmFm objects. In this paper, we use the terms Am and AmFm equivalently.

\cite{Watson1970, Watson1971} and \cite{Smith1971} suggested that atomic diffusion including the effect of radiative accelerations ($g_\mathrm{rad}$) could explain most of the observed abundances. Indeed, atomic diffusion \citep{Michaud1970} mainly involves two forces: gravity, which drags the elements towards the stellar centre, and $g_\mathrm{rad}$, which push the matter upwards. This motion is defined with respect to a buffer gas, composed of hydrogen in the envelopes of stars during most of their evolution. Radiative accelerations arise from the momentum transfer between the radiation field and the ions of the medium. Their strength for a given species thus depends on its ability to absorb photons, which vary with depth, according to the local dominant ionisation stage and abundance. Each chemical then migrates, creating local accumulations or depletions. Because of the interplay between $g_\mathrm{rad}$ and abundances, the abundance stratification process is non-linear, and time-dependent calculations are required to determine the abundance profile with depth of each species as the star evolves.

Other transport processes are also at play in the stellar medium: large-scale hydrodynamical motions interact with the abundance stratification that atomic diffusion tends to build very slowly. These processes generate turbulent mixing and tend to reduce the abundance gradients, and sometimes homogenise the medium as in convective zones. Constraints for turbulent mixing were set by \cite{Richer_etal2000} using the Montpellier-Montréal code \citep{Turcotte_etal1998b}, with the conclusion that the superficial layers of Am stars should be fully mixed down to the depth where iron-peak elements dominate the opacity (the so-called `$Z$-bump' occurring around $T=200,000~\mathrm K$). Additional mixing using a parametrised profile is also needed for the abundance pattern of Am stars to be reproduced. \cite{Talon_etal2006} investigated how meridional circulation could account for this additional mixing. Fingering mixing can also alter the chemical stratification whenever inverse mean molecular weight ($\mu$) gradients appear. These inversions can be due to atomic diffusion itself, as shown by \cite{Theado_etal2009} with the Toulouse-Geneva evolution code \citep[TGEC,][]{Hui-Bon-Hoa2008,Theado_etal2012,Hui-Bon-Hoa_Vauclair2018}. To match the observed abundances in Am stars, \cite{Deal_etal2016} showed that the different mixing zones (either fingering or dynamical convection zones) have to be connected down to the $Z$-bump, in agreement with the results of \cite{Richer_etal2000}. The role of mass loss has been studied by \cite{Vick_etal2010}, who showed that mass-loss rates between $5.10^{-14}$ and $10^{-13}M_\odot /\mathrm{yr}$ can produce surface abundances consistent with those observed in many cluster Am stars.

The computation of the $g_\mathrm{rad}$ of a given chemical element requires extensive atomic data for each of its ions in order to estimate the momentum transfer through the different types of interaction with the radiation field (bound-bound, bound-free, free-free, and scattering). These very same data are used to compute the Rosseland mean needed to build the stellar structure. Efforts have been made since the 1990s to provide extensive and consistent sets of computed atomic data for elements that contribute significantly to the mean Rosseland opacity, with the aim of improving stellar models, and in particular those of variable stars. In the abovementioned modelling of Am stars, TGEC uses the Opacity Project data \citep{Badnell_etal2005} and the Montpellier-Montréal code those of OPAL \citep{Iglesias_Rogers1996}.

Because of its negligible contribution to the Rosseland opacity, scandium is absent from both datasets. To overcome the lack of atomic data, \cite{Alecian_Artru1990a} proposed a method to obtain approximate $g_\mathrm{rad}$ by interpolation between neighbouring elements for ions having the same number of electrons (that is, along iso-electronic sequences). With this method, \cite{Alecian1996} computed $g_\mathrm{rad}$ for scandium in static Am stars envelopes, and proposed scenarios for the evolution of its surface abundance with time, depending on the thickness of the superficial mixing zone and on the mass-loss rate. \cite{LeBlanc_Alecian2008} also used interpolations for Sc due to the lack of atomic data at that time, but with the parametric form of the $g_\mathrm{rad}$ expression of \cite{Alecian_LeBlanc2002} and \cite{LeBlanc_Alecian2004} (the so-called single-valued parameter, or SVP method). Their calculations showed that the surface under-abundance of Sc could be produced at the bottom of the superficial hydrogen convection zone. However, they suggested that only detailed evolutionary calculations could confirm this.

\cite{Massacrier_Artru2012} computed a whole set of atomic data for all the scandium ions except ScI and ScII. Their data were used by \cite{Alecian_etal2013} to calculate more accurate $g_\mathrm{rad}$ for Sc in a static model, with the same interrogation as \cite{LeBlanc_Alecian2008} regarding the thickness of the surface mixing zone needed to reproduce the observed anomalies for Sc. They also insisted on the need for evolutionary models to answer this question. With the implementation of the calculation of $g_\mathrm{rad}$  with the SVP approximation in TGEC \citep{Theado_etal2009}, computations of evolutionary models including the effect of atomic diffusion for scandium are now within reach thanks to the set of SVP parameters computed by \cite{Alecian_etal2013}.

After a presentation of the TGEC code in Sect.~\ref{tgec}, we show the time evolution obtained for the surface abundance of scandium according to different scenarios for mixing and mass loss (Sect.~\ref{results}). We then compare them to observed surface abundances in AmFm stars (Sect.~\ref{observations}) and discuss the constraints brought by the abundance of scandium (Sect.~\ref{discussion}).


\section{Numerical models}\label{tgec}

\subsection{Basic input physics}

We used the \cite{Hui-Bon-Hoa_Vauclair2018} version of the TGEC code, in which higher computational performances are reached through parallelisation and adaptative diffusion time steps. Also, the numerical method has been revised to ensure a better respect of mass conservation for each chemical species.

The stellar structure is converged at each time step assuming hydrostatic equilibrium. The Rosseland mean opacities are computed on-the-fly with the method described in \cite{Hui-Bon-Hoa2021} and the monochromatic opacities from the Opacity Project (OP) OPCD v.3.3 data \citep{Seaton2005}. This ensures a full consistency between the mean opacity and the chemical composition everywhere in the star. The evolution involves nuclear reaction rates from the NACRE compilation \citep{Angulo_etal1999} and the OPAL2001 equation of state \citep{Rogers_Nayfonov2002}. Convection is treated with the mixing-length theory \citep{Bohm-Vitense1958} using a mixing-length parameter $\alpha=L/H_\mathrm{P}$ of 1.8. When full mixing down to the Z-bump is ignored, we add a mild turbulent mixing at the bottom of the superficial convective zone (SCZ) to avoid unrealistic discontinuities of the abundances \citep{Theado_etal2009}. This mixing is parametrised by a diffusion coefficient $D_\mathrm{mix}$ that writes at current radius $r$:
\begin{equation}\label{mildMixing}
D_\mathrm{mix}=D_\mathrm{bcz}\exp\left(\frac{r-r_\mathrm{bcz}}{\Delta}\ln 2\right),
\end{equation}
where $D_\mathrm{bcz}$ is the diffusion coefficient at the bottom of the SCZ (at radius $r_\mathrm{bcz}$) and  here it is set to $10^5 \mathrm{cm^2.s^{-1}}$. The depth of the mixing is expressed by $\Delta$, chosen as 0.2~\% of the stellar radius. If full mixing down to the Z-bump is considered, we refer to this fully mixed layer as the surface mixing zone (hereafter SMZ) to emphasise the difference with the mixing due to convection. We used Eq.~1 of \cite{Richer_etal2000} for the diffusion coefficient $D_\mathrm{T}$ of turbulent mixing below the SMZ:
\begin{equation}
D_\mathrm{T}=\omega D(\mathrm{He})_0\left(\frac{\rho_0}{\rho}\right)^n\label{equation:RMT},
\end{equation}
where $D(\mathrm{He})_0$ is the diffusion coefficient of helium at density $\rho_0$. In our calculations, $\rho_0=\rho(T_0),$ where $T_0$ is the temperature at the bottom of the superficial mixing zone, that is $\log T=5.3$, or slightly higher when the Fe convective zone is present. The exponent $n$ is a free parameter and is set to three as in \cite{Richard_etal2001}, in the middle of the interval of \cite{Richer_etal2000}. The models are computed with either a value of 50 or 500 for the other free parameter $\omega$, which are representative of those used in these studies. No core overshoot has been considered.

When the chemical composition is not homogeneous, the mean molecular weight $\mu$ varies from one layer to the other. If $\mu$ increases outwards (unstable $\mu$-gradient), a double-diffusive instability appears, leading to fingering mixing \citep[e.g.][ and references therein]{Vauclair2004,Zemskova_etal2014}. The effect of this hydrodynamical process on the abundances is treated as a turbulent mixing with an effective diffusion coefficient estimated according to \cite{Brown_etal2013}. The $\mu$-gradients are computed assuming a fully ionised medium and are updated at each diffusion time step according to the current chemical stratification. We assume that rotation is slow enough to neglect the mixing associated with meridional circulation.

The initial abundances are the solar ones as determined by \cite{Asplund_etal2009}. We use the meteoritic values of \cite{Lodders_Palme_Gail2009} for the refractory elements, as suggested by \cite{Serenelli2010}.

\subsection{Atomic diffusion}

Atomic diffusion is considered for 16 elements: H, He, C, N, O, Ne, Na, Mg, Al, Si, S, Ar, Sc, Ca, Fe, and Ni. Its implementation uses the \cite{Chapman_Cowling1970} formalism, in which the chemicals move with respect to the major species, namely hydrogen in the case of main-sequence (MS) stars. Hydrogen is assumed to dominate the chemical composition everywhere diffusion can have a significant effect in the MS lifetime (i.e. in the stellar envelope), the other species being considered as trace elements (test-atom approximation). For helium, we use the method of \cite{Montmerle_Michaud1976}. The diffusion coefficients are computed according to \cite{Paquette_etal1986}.

\subsection{Radiative accelerations}

\begin{figure}
   \centering
   \includegraphics[width=.5\textwidth]{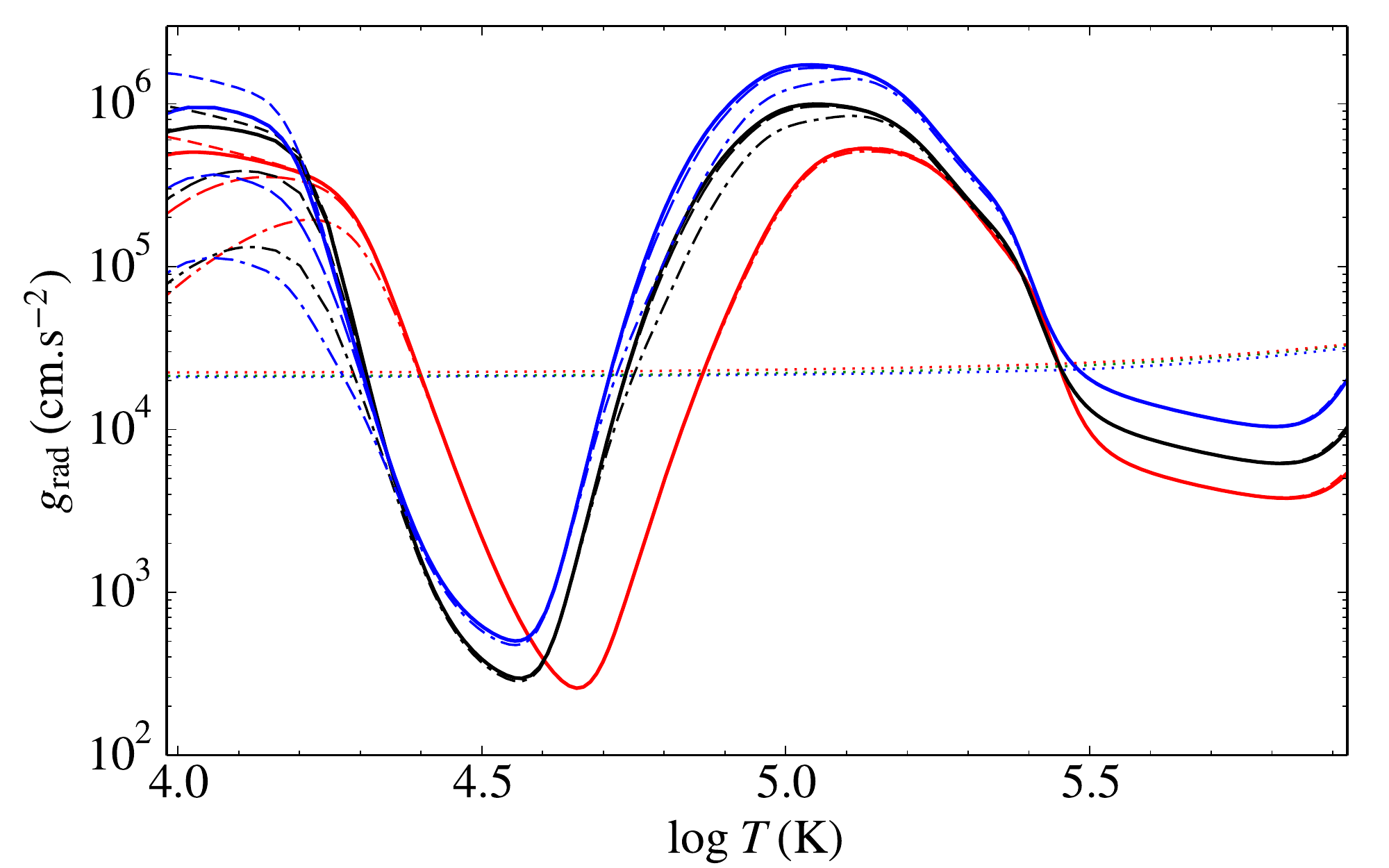}
      \caption{Radiative accelerations for Sc vs. $\log T$ at the beginning of the MS for a chemically homogeneous $1.5$, $1.7$, and $2~{M_\odot}$ model (red, black, and blue sets of curves, respectively). For each mass, the short-dashed, long-dashed, and dash-dotted curves show the $g_\mathrm{rad}$ for an abundance one-tenth, ten times, and a hundred times the solar value respectively. The dotted curves denote the local gravity.
              }
         \label{figure:g_rad}
\end{figure}

The $g_\mathrm{rad}$ are calculated at each diffusion time step with the SVP method \citep{Alecian_LeBlanc2002,LeBlanc_Alecian2004} using the current local abundances. The SVP method has the advantage of being numerically expedient compared to other methods commonly used to calculate $g_\mathrm{rad}$, such as opacity sampling \citep[e.g.][]{LeBlanc_etal2000}. In the SVP approach, the $g_\mathrm{rad}$ are computed using analytic functions, involving sets of pretabulated parameters for each element. These sets are computed for a solar composition, and for stellar masses between 1 and 5~$M_\odot$ by steps of 1~$M_\odot$. When needed, they are interpolated between the mass grid points. The values for scandium are available only for 1.5 and 2~$M_\odot$ population I stars \citep{Alecian_etal2013}, that is, the cool part of the AmFm mass range. The SVP parameters have been updated and the mass range extended to $10~M_\odot$ by \cite{Alecian_LeBlanc2020}, but scandium has not been addressed. Therefore, we use the previous dataset of \cite{LeBlanc_Alecian2004} for all the chemical elements  except for Sc where the results of \cite{Alecian_etal2013} are employed. For Ni, we use the SVP parameters derived by \cite{Hui-Bon-Hoa_Vauclair2018}. They are first extrapolated from those of Fe along iso-electronic sequences, and then fitted to the $g_\mathrm{rad}$ computed with the Montpellier-Montréal code (O. Richard, private communication). We found a good agreement for the nickel abundance and $g_\mathrm{rad}$ profiles versus depth between the two codes, at different ages of the evolution of a $1.7~{M_\odot}$ star computed with similar input parameters.\\
Figure~\ref{figure:g_rad} shows the $g_\mathrm{rad}$ of Sc versus temperature for three models of mass 1.5, 1.7, and $2~{M_\odot}$, which will be considered in this paper. The value of 1.7~${M_\odot}$ was chosen to allow direct comparisons with previous studies. These $g_\mathrm{rad}$ are computed with homogeneous solar abundances at the beginning of the MS. Our results for the 1.7, and $2~{M_\odot}$ models are very close to that of the $1.9~{M_\odot}$ model shown in Fig.~1 of \cite{Alecian_etal2013}. To quantify the effect of the scandium content, calculations made with abundances ten times lower, ten times higher, and a hundred times higher than the solar value show that its $g_\mathrm{rad}$ are quite insensitive to abundance variations in this abundance range, apart from the uppermost layers (below $\log\ T=4.3$) where they are ten times lower for a 2~dex over-abundance, whatever the model (see Fig.~\ref{figure:g_rad}). A small effect is also found around $\log\ T=5$ for the 1.7 and $2.0~{M_\odot}$ models, for which the $g_\mathrm{rad}$ are decreased by about 0.15~dex for a hundred times the solar content. So, apart from the uppermost layers, we conclude that the $g_\mathrm{rad}$ of Sc are mainly due to the absorption by spectral lines in an unsaturated regime. The weak dependence with respect to the local abundance allows us to rely on these $g_\mathrm{rad}$ variations with depth to explain the abundance variations of scandium with time, except when too important over-abundances occur.

\subsection{Mass loss}\label{massloss}

Mass loss is treated the same way as in \cite{Vick_etal2010}: the outermost layers are removed at each evolution time step, whereas the effects of nuclear reactions and atomic diffusion are computed between these time steps with an operator splitting scheme. The maximum time step is set to restrain the amount of material removed to the mass contained in the SCZ. We therefore avoid removing layers where chemical inhomogeneities have been created by atomic diffusion. The major difference with the method of \cite{Vick_etal2010} is that our spatial grid points are assigned to constant masses and are unchanged from one time step to another, so that we have no drift velocity to add to the diffusion velocity. Interpolations are made in the outermost layers to maintain spatial resolution (see the discussion in their Sect.~4.1.1). As in the bulk of their calculations, we also consider unseparated winds here, in which the amount of each element lost at each time step only depends on its mass fraction in the layers that are removed. The mass-loss rates we chose are in the same interval as those of their $1.5~M_\odot$ model, omitting their strongest value ($10^{-12}~M_\odot/\mathrm {yr}$) since it eliminates any chemical separation. The three values we adopted ($10^{-13}$, $3.1\times10^{-14}$, and $10^{-14}~M_\odot/\mathrm {yr}$) have a constant step on a logarithmic scale.

\begin{figure}
   \centering
   \includegraphics[width=.5\textwidth]{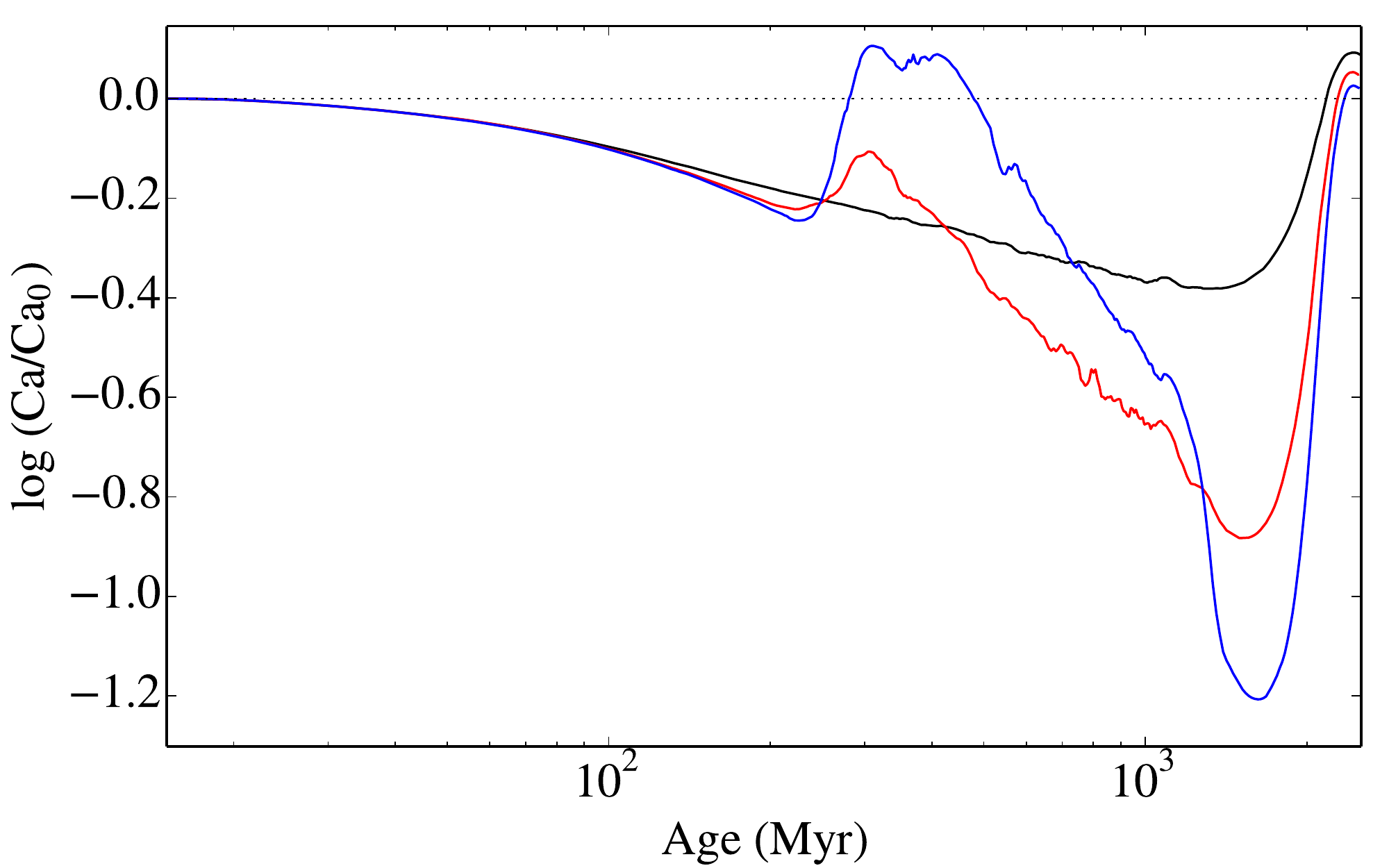}
      \caption{Evolution with time of the superficial abundance of Ca for the $1.5~M_\odot$ model with various mass-loss rates ($10^{-13}$, $3.1\times10^{-14}$, and $10^{-14}~M_\odot/\mathrm {yr}$, shown by the black, red, and blue curves, respectively). The abundances are expressed as the logarithm of the ratio of the current mass fraction and the initial one. The thin dotted curve represents the initial abundance.
              }
         \label{figure:compareML}
\end{figure}

To compare the two implementations, we rely on the upper panel of \cite{Vick_etal2010}'s Fig.~11, which shows the time evolution of the surface abundance for several elements in a 1.5~$M_\odot$ model. For the sake of conciseness, among all the metals of this figure in common in the two studies (namely O, S, Ca, Fe, and Ni), we only detail the case of calcium since it is another key element to the Am phenomenon: just as for scandium, under-abundances of calcium are typical of Am stars. Figure~\ref{figure:compareML} shows the time evolution of the mass fraction of Ca in our $1.5~M_\odot$ model. In the comparison of our $10^{-13}~M_\odot/\mathrm {yr}$ model with the model of same mass-loss rate in the upper panel of Fig.~11 of \cite{Vick_etal2010}, we obtained the same order of magnitude for the minimum value of the Ca surface abundance, which occurs at roughly the same age in the two studies. The minimum of our $3.1\times10^{-14}~M_\odot/\mathrm {yr}$ model is between their $2\times 10^{-14}~M_\odot/\mathrm {yr}$ and their $5\times 10^{-14}~M_\odot/\mathrm {yr}$ models minima, and is met at a similar age. We could not compare the minimum abundance of the $10^{-14}~M_\odot/\mathrm {yr}$ models because their calculation stopped at 550~Myr. The time evolution of the superficial calcium abundance in our models is somewhat different at younger ages: the amount of Ca decreases first before experiencing a peak, whereas the \cite{Vick_etal2010} models show a peak at the very beginning of the evolution, whatever the mass-loss rate. This discrepancy is due to a much thicker surface convection zone in our models at the beginning of the evolution. On a scale expressed with $\log (\Delta M/M_*)$ where $\Delta M$ is the mass above the layer under consideration and $M_*$ the mass of the star, the bottom of our SCZ lies below $\log (\Delta M/M_*)=-6$ for our models, whatever the mass-loss rate, as compared with $\log (\Delta M/M_*)\simeq-7.5$ in theirs. This deeper SCZ has its bottom around the dip of Ca abundance, near $\log\ T=5.3$, explaining why the Ca superficial abundance in our model begins with a depletion phase. The difference in the extent of the convective zone can be due to a different opacity dataset used here, a different equation of state, a different $T-\tau$ relation in the atmosphere, and a different mixing length, which was set at 2.096 in \cite{Vick_etal2010} through a solar calibration using a \cite{Krishna-Swamy1966} outer boundary condition. We therefore suspect that the differences between the results of the two studies come from these differences rather than from the detail of the implementation of mass loss. The comparison with the other chemical elements presented in their Fig.~11 also shows a good overall agreement, but with some discrepancies during the early phases of the abundance evolution due to a different thickness of the SCZ between the two models.

\section{Results}\label{results}

 We present the time evolution of the surface abundance of Sc for 1.5, 1.7, and 2~${M_\odot}$ models. These evolutionary models span from the pre-main sequence (PMS) to core hydrogen exhaustion. Atomic diffusion is considered all along the evolution, but has visible effects only once the PMS SCZ has become thin enough. Therefore, depending on the stellar mass, the abundances may not be homogeneous in the envelope at the beginning of the MS. Since we aim to complement the results of previous studies that involved other chemical species \citep[e.g.][]{Turcotte_etal1998b,Richer_etal2000,Vick_etal2010,Theado_etal2012,Deal_etal2016}, we only consider here the physical processes that have already been invoked in these studies. An investigation of the effect of new processes should not be restricted to scandium and would require all the available chemical species to be considered. This is out of the scope of this paper. 

\subsection{Models without mass loss}

For each mass, the evolution is computed with the same set of three different assumptions for the mixing processes: convection only, convection and fingering mixing, and full mixing down to the Z-bump (models labelled RMT$\omega$, with $\omega=50$ or $500$, $\omega$ being defined in Eq.~\ref{equation:RMT}).

\begin{figure}
   \centering
    \includegraphics[width=.5\textwidth]{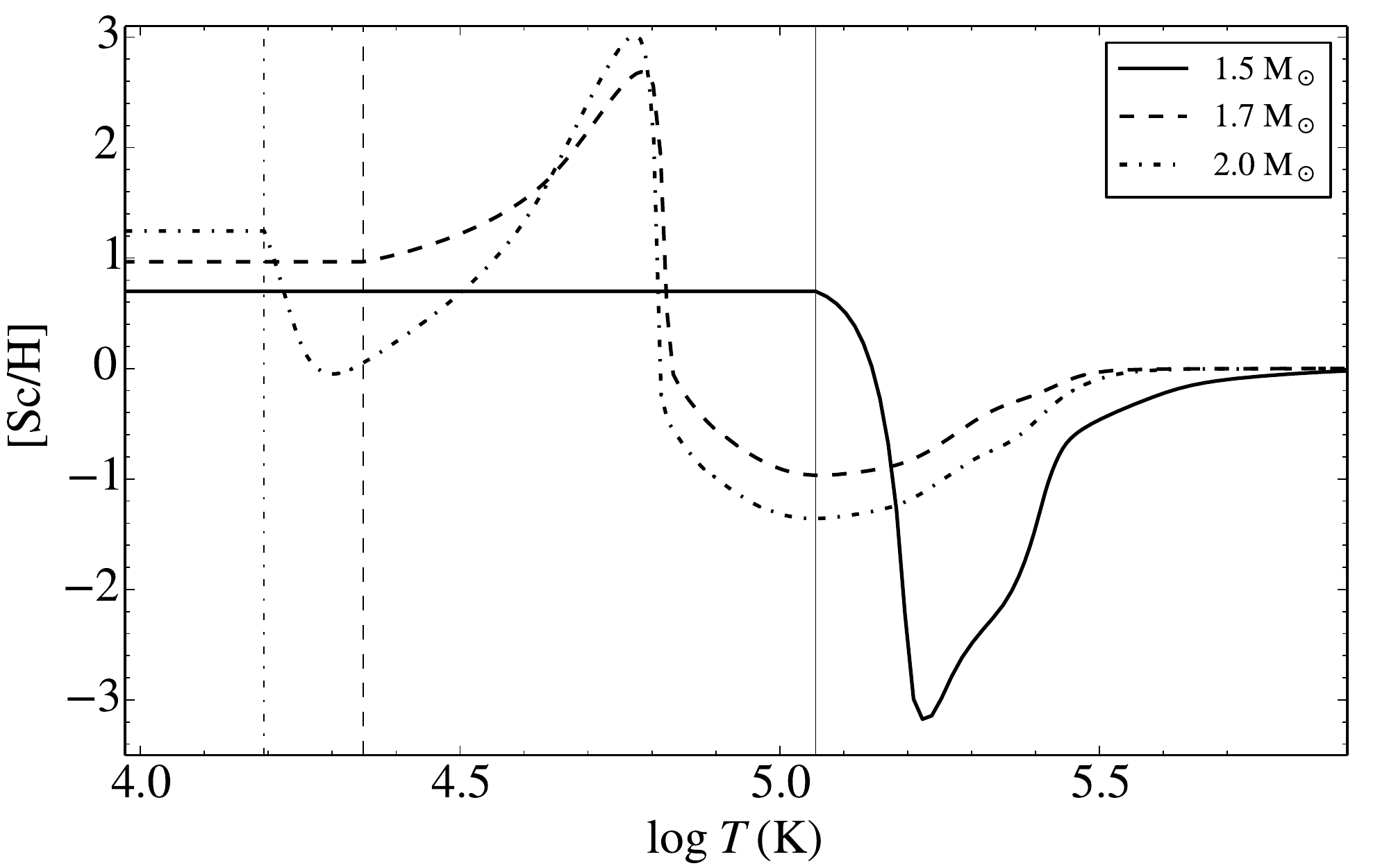}
      \caption{Abundance of Sc vs. $\log\ T$ at the time the SCZ becomes stable for the 1.5 (solid curve), 1.7 (dashed curve), and 2~${M_\odot}$ (dash-dotted curve) models, expressed as the difference between the logarithm of the current ratio to H in number fraction and the corresponding initial value. The vertical lines with the same curve style denote the location of the bottom of the surface convective zone. These results are for models with convection as the unique mixing process and with no mass loss.
              }
         \label{figure:abond}
\end{figure}

Figure~\ref{figure:abond} shows the abundance variations of Sc versus temperature for the three models when only convection is considered and at an age for which the location of the bottom of the surface convective zone has become stable, that is, when the He contribution to the surface convection zone has disappeared due to the gravitational settling of He. This happens at the beginning of the MS for the $1.7$ and $2.0~M_\odot$ models, and at about 320~Myr for the $1.5~M_\odot$ model. Strictly speaking, these profiles are only valid for the models with convection as the unique mixing process, but they will nevertheless be useful to understand the behaviour of the scandium abundance in models with other mixing prescriptions.

The abundance variation with depth is due to the corresponding variation of the $g_\mathrm{rad}$. The depletion zones around $\log\ T=5.3$ in the 1.5~${M_\odot}$ model and around $\log\ T=5.05$ in the 1.7 and 2.0~${M_\odot}$ models are related to the decrease in $g_\mathrm{rad}$ for $\log\ T$ increasing from 5.2 to 5.8. For the 1.7 and $2.0~{M_\odot}$ models, the accumulation zone around $\log\ T=4.75$ is due to the increase in the radiative acceleration from $\log\ T=4.6$ to $\log\ T=5.0$. This feature of $g_\mathrm{rad}$ sits slightly deeper in the 1.5~${M_\odot}$ model and is responsible of the higher abundance at the upper edge than at the lower edge of the depletion zone, causing a surface over-abundance. A second abundance dip appears near the surface in the $2.0~{M_\odot}$ model at $\log\ T\simeq4.3$, due to the increase in $g_\mathrm{rad}$ with decreasing $T$ above $\log\ T=4.5$.

We now focus on the time evolution of the Sc surface abundance for the three masses (1.5, 1.7, and $2.0~M_\odot$, from the top to the bottom panel of Fig.~\ref{figure:noML}, respectively), and with different assumptions for the mixing processes. No mass loss is considered in this section. Convection alone is shown by the black curves, and convection with fingering mixing by the red curves. The full mixing down to the Z-bump has been computed with two values for $\omega$ (50 and 500, represented by the blue and green curves respectively).\\

\subsubsection{The $\mathsfit{1.5~{M_\odot}}$ models}

\begin{figure}
   \centering
    \includegraphics[width=.5\textwidth]{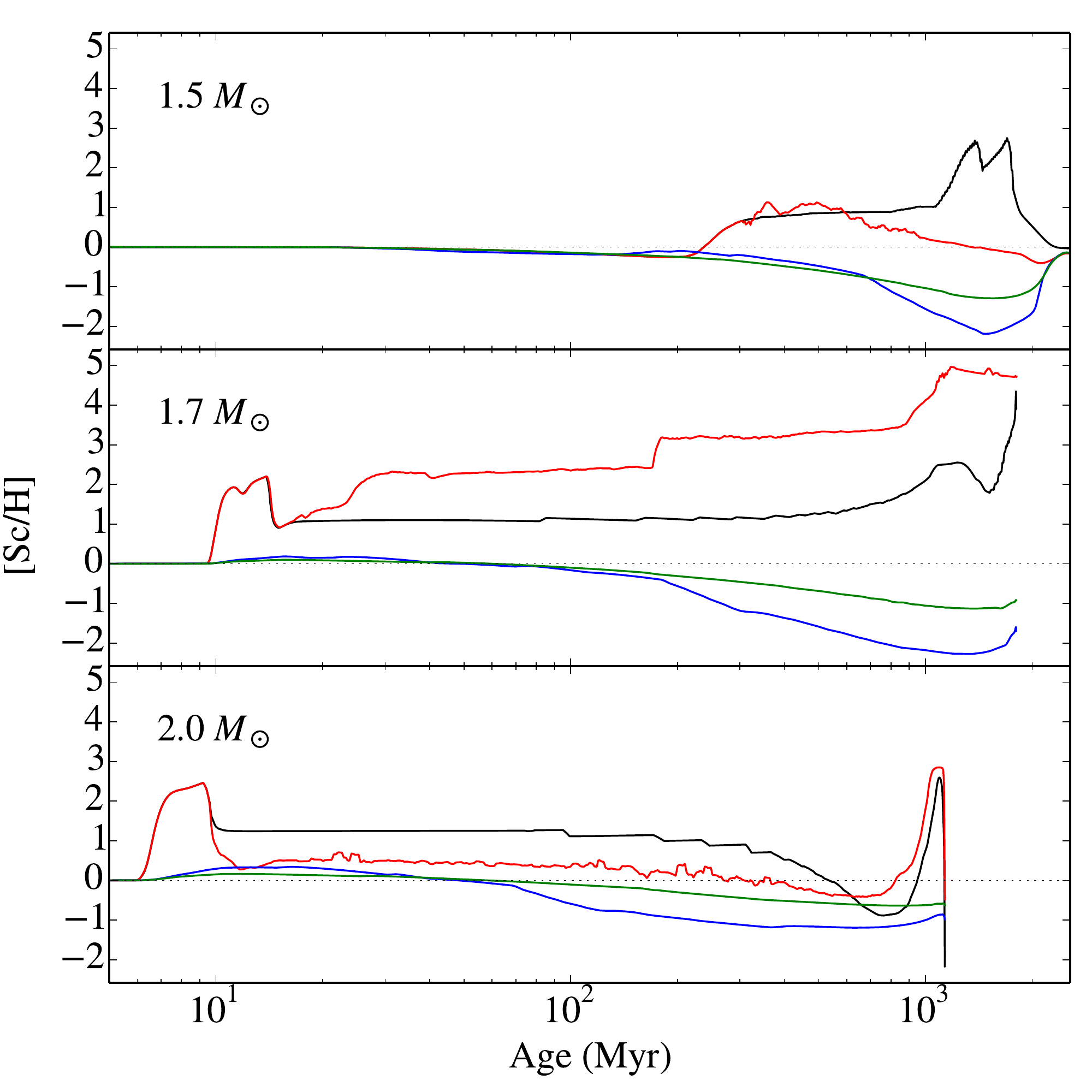}
      \caption{Evolution with time of the superficial abundance of Sc for models with convection only (black curves), with convection and fingering mixing (red curves), and with full mixing down to the Z-bump with $\omega=50$ (blue curves) and 500 (green curves). The stellar mass is shown is each panel. The thin dotted line represents the initial abundance. The vertical axis has the same range in all the panels. The thin dotted curve represents the initial abundance.
              }
         \label{figure:noML}
\end{figure}

When only convective mixing is considered, the Sc abundance first slightly decreases because the bottom of the SCZ is located in a region where Sc is depleted, around $\log T=5.3$ (black curve in the upper panel of Fig.~\ref{figure:noML}). Afterwards, between 200 and 300~Myr, the SCZ becomes thinner because of the settling of He and its bottom reaches an over-abundant region linked to a decreasing $g_\mathrm{rad}$ with increasing radius around $\log T= 5.1$, which leads to an over-abundance in the SCZ. Past 1~Gyr, the abundance of Sc increases quickly due to several oscillations of the base of the SCZ in a region where the Sc abundance gradient has become very steep. Each time the SCZ recedes, a rapid accumulation of Sc occurs due to a smaller diffusion time scale at smaller densities, which is thereafter mixed in the convective zone when it deepens again. The dip between the two peaks is related to a temporary connection between the H and He convection zone with that of the Fe-peak elements. No other connection occurs before the end of the MS life, where the thickening of the SCZ gradually erases the departure from the initial surface abundance. At the very end of the evolution, in the core contraction phase, the SCZ becomes thinner in a time scale much smaller than that of diffusion, so that no visible abundance change appears.

When fingering mixing is included in the calculations, the surface Sc abundance evolution with time is the same as in the previous case at the beginning of the evolution. This similarity is due to the gravitational settling of helium, whose abundance gradient stabilises the medium against $\mu$-gradient inversions. When the He abundance is low enough in the superficial layers, from 300~Myr on, its stabilising ability is weakened, and inverse $\mu$-gradients appear from place to place due to the departures from the initial abundances of the various chemicals from one layer to the other. Fingering mixing acts first just below the SCZ and changes slightly its thickness, leading to some oscillation of the Sc surface abundance compared to the track of the model with convective mixing only. As time goes by, deeper layers experience fingering mixing until the depletion zone around $\log\ T=5.3$ is reached so that the Sc surface abundance decreases with time. No peak appears near the end of the MS phase because the superficial H and He convective zone is connected to that of the iron-peak elements at 1~Gyr.

The models with full mixing down to the Z-bump exhibit surface Sc under-abundances all along the evolution because the bottom of the SMZ is in layers where Sc is depleted. For the RMT50 model, the Sc surface abundance track departs from a strictly monotonous decrease between 100 and 300~Myr because a slight recession of the SMZ in this time interval shifts its bottom towards layers with greater $g_\mathrm{rad}$. The discontinuity of the slope around 600~Myr is due to the appearance of the iron-peak convective zone, which merges immediately with the former SMZ. The time variation of the scandium surface abundance is smoother in the RMT500 model because the mixing is strong enough below the SMZ to flatten the abundance variation with depth, so that the He convection zone persists all along the evolution. Besides, the SMZ merges progressively with the iron-peak element convective zone so that the change of slope of the abundance track is less steep.

\subsubsection{The $\mathsfit{1.7~{M_\odot}}$ models}

At the beginning of the evolution of the convection mixing model (black curve in the middle panel of Fig.~\ref{figure:noML}), the bottom of the SCZ is located around $\log T=4.7$, which is near the maximum of the accumulation peak shown in Fig.~\ref{figure:abond}. Later on, the He abundance becomes low enough for the He convective zone to disappear and the bottom of the SCZ is shifted to $\log\ T=4.35$, where the Sc over-abundances are smaller. The scandium surface abundance later increases gradually with the thickening of the SCZ during the stellar evolution, with layers richer in Sc being progressively merged with the SCZ. This process speeds up at the end of the MS, and oscillations of the location of the bottom of the SCZ in layers where the Sc abundance gradient is strong leads to the strong increase at the end of the evolution, especially when the SCZ recedes during the core contraction phase. To test the influence of the initial chemical composition, a model has been computed using the \cite{Grevesse_Noels1993} abundances, and we obtained very similar results (not shown) as for the \cite{Asplund_etal2009} abundances.

When fingering mixing is considered, inverse $\mu$-gradients appear only once the He convective zone has disappeared, so that the beginning of the surface abundance evolution track is the same as in the previous calculation. Afterwards, fingering mixing occurs because of the iron accumulation below the SCZ, so that the slab of envelope where Sc is over-abundant extends progressively down to $\log\ T=5.5$. Below this layer, $g_\mathrm{rad}$ is lower than the local gravity, thus setting a limit to the amount of scandium in its over-abundance peak. The surface abundance is then almost constant between 30 and 180~Myr, until the zone mixed through fingering mixing extends towards the maximum of the scandium abundance peak, yielding a sudden increase in the Sc superficial abundance. This fingering mixing zone remains until it gradually merges with the SCZ at the end of the MS, causing a large increase in the Sc surface abundance around 1~Gyr.

With full mixing down to the Z-bump, the surface scandium abundance track has a similar shape as for the $1.5~{M_\odot}$ model with a shift in age. A noticeable difference lies in a slight and transient over-abundance near the beginning of our modelling due to a thinner SMZ with a bottom located at $\log\ T=5.3$. At this location, $g_\mathrm{rad}$ is greater than the local gravity, and scandium accumulates in the SMZ. But simultaneously, deeper layers are slowly depleted, leading to a decrease in the Sc abundance in the SMZ after 25~Myr.

\subsubsection{The $\mathsfit{2~{M_\odot}}$ models}

For the same reason as for the $1.7~{M_\odot}$ convection model, the time evolution of the Sc abundance for the $2~{M_\odot}$ convection model begins with a peak.
However, after this initial peak, the SCZ is much thinner when the He contribution disappears compared to the $1.7~{M_\odot}$ model with a bottom around $\log\ T=4.2$. When its base deepens during evolution, more depleted layers are merged into the SCZ, leading to the progressive Sc abundance decrease with time. The peak at the end of the MS occurs when the bottom of the SCZ reaches the enriched layers near the abundance maximum around $\log\ T=4.7$.

As for the $1.7~{M_\odot}$ model, fingering mixing is at play once the SCZ settles its base around $\log\ T=4.2$, when the He convective zone has disappeared, so that the beginning of the Sc surface abundance track is similar to that with convection only. The extent and the location of the various fingering mixing zones then remain roughly constant, yielding an almost constant Sc surface abundance. Its slow decrease is due to the depletion of the layers between the surface and the point where $g_\mathrm{rad}$ becomes lower than gravity at around $\log\ T=5.5$. The surface abundance peak near the end of the MS is caused by the thickening of the SCZ, whose bottom gets gradually closer to the maximum of the Sc abundance with depth around $\log\ T=4.7$.

When the superficial layers are mixed down to the Z-bump, the evolution of the Sc surface abundance is very similar to that of the $1.7~{M_\odot}$ model, except that the initial over-abundance is greater because of a higher $g_\mathrm{rad}$ for the present stellar mass. The abundance decrease with time is also slower due to this higher $g_\mathrm{rad}$.

\subsection{Models with mass loss}

\begin{figure}
   \centering
   \includegraphics[width=.5\textwidth]{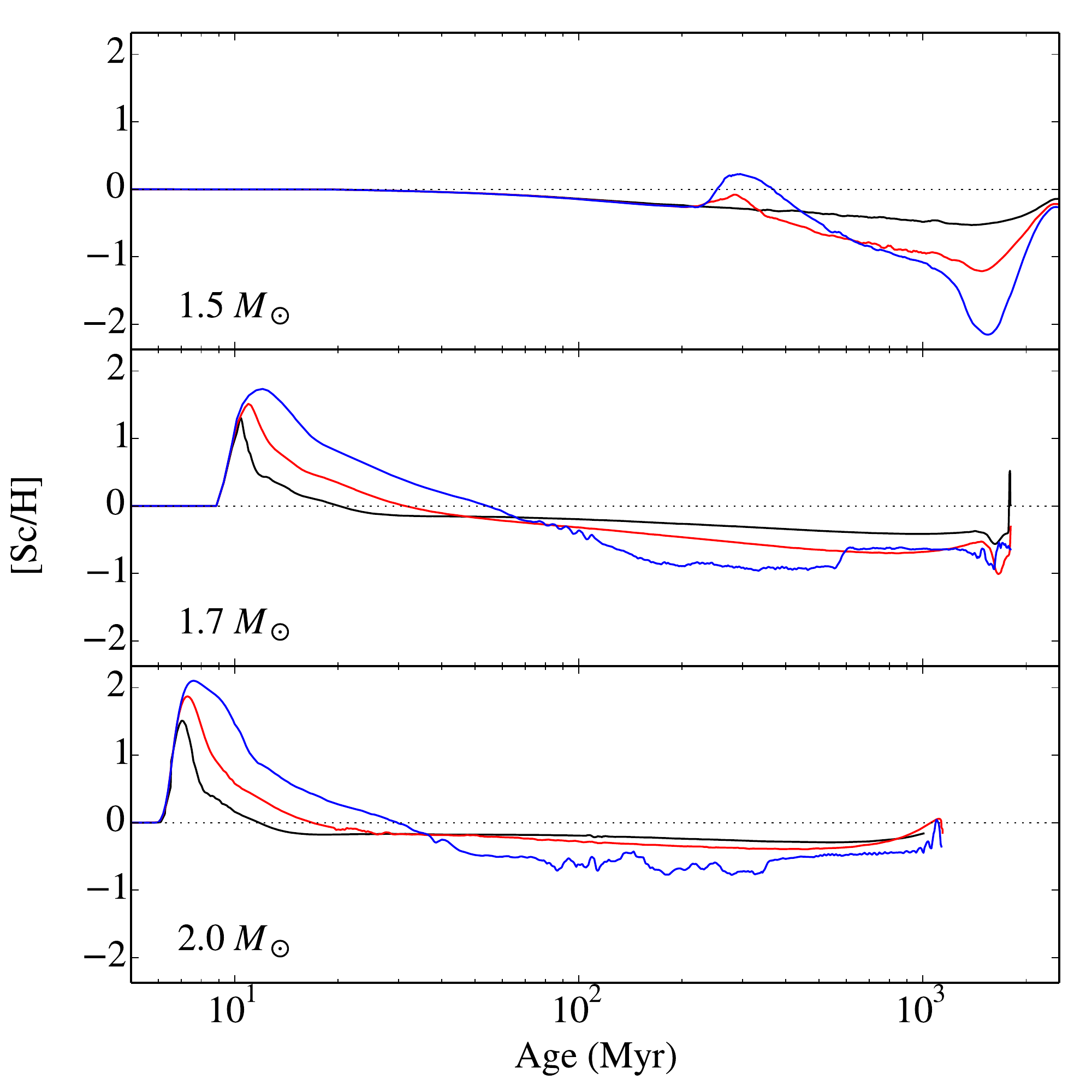}
     \caption{Same as Fig.~\ref{figure:noML} but for models with mass loss of rate $10^{-13}$ (black curves), $3.1\times10^{-14}$ (red curves), and $10^{-14}~M_\odot/\mathrm {yr}$ (blue curves).
              }
         \label{figure:ML}
\end{figure}

In this section, we consider models with mass loss. The evolution of the surface Sc abundance is shown in Fig.~\ref{figure:ML} for the three mass-loss rates considered, namely $10^{-13}$ (black curves), $3.1\times10^{-14}$ (red curves), and $10^{-14}~M_\odot/\mathrm {yr}$ (blue curves), and for the three model masses (1.5, 1.7, and $2.0~M_\odot$, from the top to the bottom panel, respectively). Convective mixing is considered, along with a mild turbulent mixing below the SCZ as defined in Eq.~\ref{mildMixing}.

The Sc surface abundance decreases slowly with time at the beginning of the evolution of the $1.5~M_\odot$ model due to the depletion around $\log\ T=5.3$. An abundance bump occurs when the He convection zone disappears setting the bottom of the SCZ around $\log\ T=5.0$, which induces an accumulation below it. But then, since the amount of Sc available is limited by the change of sign of $g_\mathrm{rad}$ at $\log\ T=5.5$, as mass is removed, the Sc content diminishes in the outer envelope, yielding under-abundances. This bump is absent with a $10^{-13}~M_\odot/\mathrm {yr}$ mass-loss rate model because He does not have enough time to settle, due to the effect of the removal of the outer layers on the stellar structure.

Whatever the mass-loss rate, the scandium abundance time evolution is close in shape to that of Ca shown in Fig.~\ref{figure:compareML}, apart from more important under-abundances. This similarity is due to similar shapes of their $g_\mathrm{rad}$ versus depth where chemical separation occurs, that is, below the bottom of the superficial mixing zone. The only difference is the location of the $g_\mathrm{rad}$ maximum, that of Ca being slightly shallower and with a less steep slope around it than for Sc. This explains the more important under-abundances of Sc for the lowest mass-loss rates, with the thickness of the SCZ decreasing with decreasing rate.

For the two other masses, their abundance tracks versus age begin with an over-abundant phase due to the accumulation below the SCZ. The lower the mass-loss rate, the longer the duration of this phase: when the rate is low, atomic diffusion has time to produce stronger departures from the initial abundance and the amount removed by mass loss is lower at each time step, allowing the bump to last longer. But in any case, under-abundances appear sooner or later as the star evolves due to the limited Sc content.


\section{Observational constraints}\label{observations}

To compare our models to observed abundances, we turned to cluster and moving group stars, whose initial chemical compositions are assumed to be identical within the cluster or group and whose ages are known with relative accuracy. We first selected a set of AmFm stars for which Sc abundances have been measured. Among such stars, we retained those for which our models are relevant: their effective temperature should be closer than the $T_\mathrm{eff}$ uncertainty to that of our models at their age. This uncertainty amounts to $\pm $200~K in the temperature range of our models \citep{Napiwotzki_etal1993}. The resulting set of stars is detailed in Table~\ref{table:observations}, along with their host cluster, age, fundamental parameters, and scandium abundance difference with respect to the initial value of the cluster, assumed to be equal to the mean abundance of normal F stars \citep[e.g.][]{Gebran_etal2008}. If not provided, we computed this mean from the data. For stars concerned by more than one study, we used average $T_\mathrm{eff}$, $\log\ g$, and abundance values. The uncertainty on the abundance takes this dispersion into account besides the systematic uncertainties. The scatter of the initial abundance is also considered.

\begin{table*}
\begin{center}
\tiny{
\begin{tabular}{c|llccccccc}
\hline
Model ($M_\odot$)&Name&Cluster&Age (Myr)&$T_\mathrm{eff}$ (K)&$\log\ g$ (cgs)&$\Delta$log(Sc/H)&$\sigma$($\Delta$log(Sc/H))&References\\
\hline
1.5&GSC 07380-01211&NGC 6405&75.0$^a$&6850.0&4.20&0.07&0.31&8\\
\hline
{\multirow{10}{*}{1.7}}
&BD+49$^\circ$967&$\alpha$ Per   &98.0$^b$&8000.0&4.15&-1.00&0.30&6\\
&HD 23325&Pleiades&100.0$^c$&7654.0&4.24&-0.44&0.31&4,7\\
&HD 107513&Coma Ber&580.0$^e$&7279.0&4.02&-0.27&0.17&3\\
&HD 27628&{\multirow{6}{*}{$\left.\vphantom{\begin{tabular}{l}1\\2\\3\\4\\5\\6\end{tabular}}\right\}$ Hyades}}&{\multirow{6}{*}{625.0$^f$}}&7315.5&4.12&-1.23&0.36&5,10\\
&HD 27749&&&7570.0&4.27&-1.06&0.30&7\\
&HD 28226&&&7465.0&4.09&-0.94&0.52&5,10\\
&HD 28546&&&7788.3&4.22&-0.55&0.30&5,7,10\\
&HD 29499&&&7751.0&4.10&0.24&0.35&10\\
&HD 33204&&&7646.0&4.11&-0.64&0.35&10\\
&HD 73045&{\multirow{2}{*}{$\left.\vphantom{\begin{tabular}{l}1\\2\end{tabular}}\right\}$ Praesepe}}&{\multirow{2}{*}{800.0$^g$}}&7545.0&4.16&-0.70&0.48&1,6\\
&HD 73818&&&7232.0&3.82&-1.93&0.30&1\\
\hline
{\multirow{20}{*}{2.0}}
&HD 318091&NGC 6405&75.0&8700.0&4.00&-0.59&0.31&8\\
&HD 6116A&{\multirow{2}{*}{$\left.\vphantom{\begin{tabular}{l}l\\l\end{tabular}}\right\}$ UMa moving gr.}}&{\multirow{2}{*}{500.0$^d$}}&8073.0&3.93&-0.58&0.22&9\\
&HD 116657&&&8425.0&4.40&-0.67&0.22&9\\
&HD 106887&{\multirow{7}{*}{$\left.\vphantom{\begin{tabular}{l}1\\2\\3\\4\\5\\6\\7\end{tabular}}\right\}$ Coma Ber}}&{\multirow{7}{*}{580.0}}&8291.0&4.20&0.01&0.37&3\\
&HD 107168&&&8291.5&4.19&-0.20&0.24&3,7\\
&HD 107276&&&8000.0&4.00&-0.08&0.27&3\\
&HD 108486&&&8172.0&4.14&-0.65&0.37&3,7\\
&HD 108642&&&8079.0&4.06&-1.19&0.24&3\\
&HD 108651&&&8090.0&4.24&-0.56&0.30&6\\
&HD 109307&&&8396.0&4.10&0.08&0.18&3\\
&HD 28355&{\multirow{3}{*}{$\left.\vphantom{\begin{tabular}{l}1\\2\\3\end{tabular}}\right\}$ Hyades}}&{\multirow{3}{*}{625.0}}&7955.0&3.96&-1.31&0.46&5,10\\
&HD 30210&&&8093.7&3.95&-1.43&0.50&5,7,10\\
&HD 33254&&&7860.0&4.16&-1.45&0.30&7\\
&HD 73045&{\multirow{6}{*}{$\left.\vphantom{\begin{tabular}{l}1\\2\\3\\4\\5\\6\end{tabular}}\right\}$ Praesepe}}&{\multirow{6}{*}{800.0}}&7545.0&4.16&-0.70&0.48&1,6\\
&HD 73618&&&8115.0&3.94&-1.10&0.86&1,6\\
&HD 73709&&&8065.0&3.86&-1.08&0.46&1,6\\
&HD 73711&&&8020.0&3.69&-1.09&0.30&1\\
&HD 73730&&&8045.0&3.96&-1.20&0.72&1,7\\
&HD 74656&&&7800.0&3.99&-0.30&0.30&2\\
\hline
\hline
\end{tabular}
}
\end{center}
\caption{Observed stars corresponding to the models. $\Delta$log(Sc/H) is the difference between the logarithm of number ratio $n_\mathrm{Sc/}n_\mathrm{H}$ and the mean value of the cluster or group. HD~73045 is compatible with both the 1.7 and the 2.0~$M_\odot$. The ages are from: $^a$\cite{Kilicoglu_etal2016}, $^b$\cite{Ortega_etal2022}, $^c$\cite{Dahm2015}, $^d$\cite{King_etal2003}, $^e$\cite{Delorme_etal2011}, $^f$\cite{Perryman_etal1998}, $^g$\cite{Gonzalez-Garcia_etal2006}.\newline 
References: $^1$\cite{Fossati_etal2007}, $^2$\cite{Fossati_etal2008}, $^3$\cite{Gebran_etal2008}, $^4$\cite{Gebran_Monier2008}, $^5$\cite{Gebran_etal2010}, $^6$\cite{Hui-Bon-Hoa_etal1997}, $^7$\cite{Hui-Bon-Hoa_Alecian1998}, $^8$\cite{Kilicoglu_etal2016}, $^9$\cite{Monier2005}, $^{10}$\cite{Varenne_Monier1999}. 
}\label{table:observations}
\end{table*}

\begin{figure}
   \centering
   \includegraphics[width=.5\textwidth]{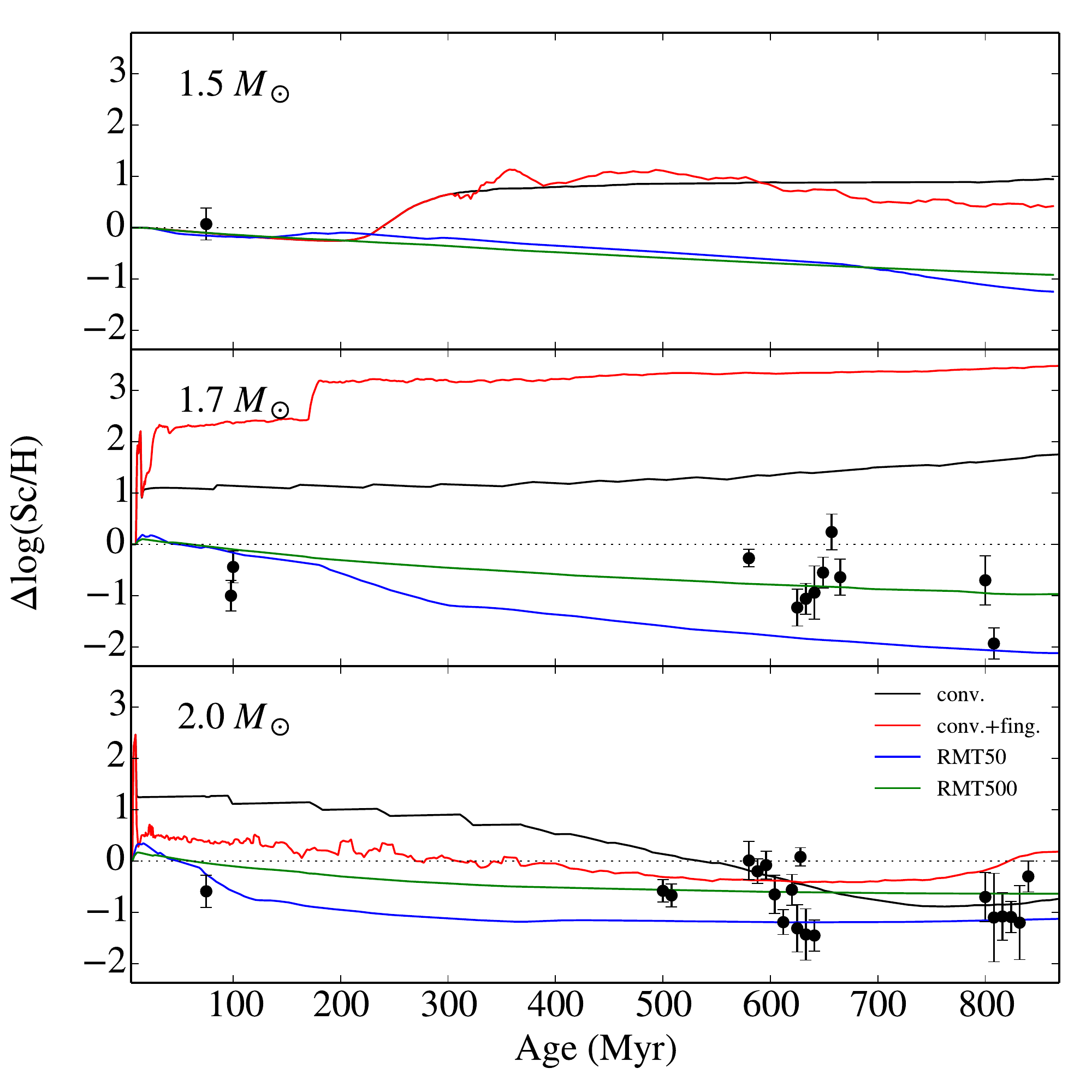}
     \caption{Comparison between the scandium surface abundance of our models without mass loss and the observed values of Table~\ref{table:observations}. The abundances are expressed as the difference between the logarithm of the actual value relative to H and the initial one. The colours of the tracks refer to the same transport process assumptions as in Fig.~\ref{figure:noML} and are recalled in the lower panel. For the sake of legibility, an offset has been added to the abscissa for stars of same age. The horizontal thin dotted line represents the initial abundance.
              }
         \label{figure:AbSurfNum}
\end{figure}

The comparison with our models without mass loss is shown in Fig.~\ref{figure:AbSurfNum}, where the time variations of the Sc surface abundance of the models are plotted along with the observed values and their uncertainties.  GSC~07380-01211 is the only star that can be reproduced by the $1.5~M_\odot$ models, and its scandium abundance is in agreement with the value we obtained whichever the prescription we considered. For the $1.7~M_\odot$ model, the convection with fingering mixing scenario fails to reproduce the observations. Considering deep mixing, the model with the RMT500 prescription is consistent with the greatest number of observed values. The RMT50 model is compatible with two stars, and the presence of objects lying between the tracks of these two models suggests that intermediate values of $\omega$ could also be considered. The picture is much less conclusive for the $2.0~M_\odot$ model as no mixing prescription can be completely ruled out. Before 500~Myr, the convection model seems the less appropriate. The youngest star is closest to the RMT50 model. Observations of other objects younger than 500~Myr could help to clarify the picture. Nevertheless, as a radical change in the nature of the transport processes is unlikely in such a narrow mass range, the models with deep mixing seem to be the most suitable.

\begin{figure}
   \centering
   \includegraphics[width=.5\textwidth]{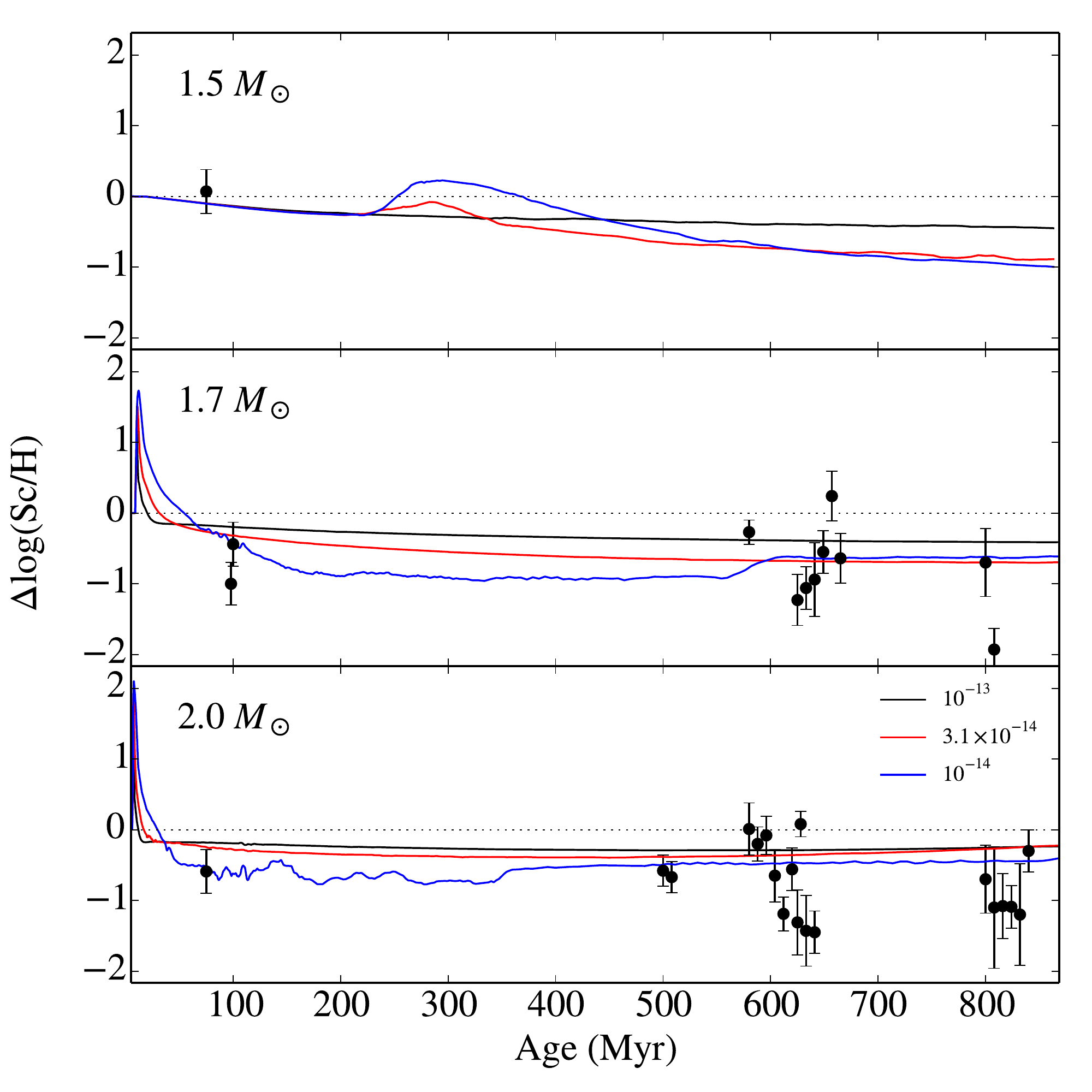}
     \caption{Same as for Fig.~\ref{figure:AbSurfNum}, but for the mass-loss models, whose rates in $M_\odot/\mathrm {yr}$ are recalled in the lower panel.
              }
         \label{figure:AbSurfNumML}
\end{figure}

Figure~\ref{figure:AbSurfNumML} shows the comparison of our models including mass loss. The $1.5~M_\odot$ model can reproduce the observed abundance whatever the mass-loss rate. All the mass-loss rates of the $1.7~M_\odot$ model are also compatible with the observations. Some stars showing a more under-abundant Sc compared to the models suggest that mass loss could be weaker than those used in our models. The same conclusion holds for the $2.0~M_\odot$ models.


\section{Discussion and conclusions}\label{discussion}

Models including atomic diffusion alone lead to surface abundances too different from those observed in AmFm stars. To reconcile models and observations, two approaches have been investigated so far: some ad hoc turbulent mixing \citep{Richer_etal2000,Deal_etal2016} or mass loss \citep{Vick_etal2010}. However, as stated by \cite{Vick_etal2010} and \cite{Michaud_etal2011b}, considering surface abundances does not allow firm conclusions to be drawn about the transport process at play, turbulence or mass loss, at least for the chemical species included in OPAL. 

We used the two above-mentioned approaches to model the evolution of the Sc surface abundance, looking for additional constraints on the transport processes. It appears that the results for scandium are globally in line with those of previous studies for the deep turbulent mixing models. We have adopted here an anchoring of the turbulent diffusion coeffcient according to temperature, as suggested by \cite{Richer_etal2000} for models that develop an iron-peak element convective zone. \cite{Michaud_etal2005} did the same to analyse the $o$ Leo system. However, other kinds of reference point have been considered in the Montr\'eal code, such as density \citep[e.g.][]{Richer_etal2000,Vick_etal2010} and mass \citep{Michaud_etal2011b}. Comparisons with individual star abundances did not allow us to conclude that a given approach was better than the others. To find empirically the most suitable anchoring method, if any, a further study could use the time variation of the abundances of cluster stars.

For the mass-loss models, some differences exist in the detailed behaviour of the abundance evolution between the Montr\'eal model and ours, especially for the mass-loss models. They can originate from the physics and data used (equation of state, opacity dataset, radiative accelerations) and the implementation of the physical processes. The whole set of chemical species in common should be considered for a detailed comparison between the two codes. Lower mass-loss rates could be considered to reproduce the Sc abundances, but the agreement with observations for the other elements has to be checked.

Still, for the turbulence as well as the mass-loss models, we would be left with several ranges for the free parameters owing to the scatter in the observed abundances, as we assume that the stars are non-rotating and without a magnetic field, and we ignored the effect of multiplicity. All these stellar properties are likely to interact with the migration of chemicals, and thus account for the surface abundance scatter. The effect of mass loss with fingering mixing as in \cite{Hui-Bon-Hoa_Vauclair2018} could also be studied. Besides, there is no reason for the physical processes in the turbulence models to be absent in cases of mass loss, and a combination of both should be investigated. In addition, it would be worth computing SVP parameters for scandium for more massive stars to encompass the hot Am stars.

\begin{acknowledgements}
We would like to thank the anonymous referee, whose questions and comments helped to improve the manuscript. This work was supported by the "Programme National de Physique Stellaire" (PNPS) of CNRS/INSU co-funded by CEA and CNES. This research was also partially funded by NSERC.
\end{acknowledgements}

%
%

   \bibliographystyle{aa} 
   \bibliography{ScAmFm}

\end{document}